\documentclass[12pt]{article}

\usepackage{amsmath,amssymb,euscript}
\usepackage{graphicx}

\makeatletter
 
  \@addtoreset{equation}{section}
 \makeatother

\newcommand{\be}{\begin{equation}}
\newcommand{\ee}{\end{equation}}
\newcommand{\bea}{\begin{eqnarray}}
\newcommand{\eea}{\end{eqnarray}}
\newcommand{\beann}{\begin{eqnarray*}}
\newcommand{\eeann}{\end{eqnarray*}}

\newcommand{\ba}{\begin{array}}
\newcommand{\ea}{\end{array}}

\newcommand{\Tr}{{\rm Tr}\,}
\newcommand{\del}{\partial}
\newcommand{\Z}{\mathbb{Z}}

\def\CC{{\cal C}}

\newcommand{\diag}{{\rm diag}\,}
\newcommand{\combi}[2]{{}_{#1}C_{#2}}

\def\N{{\cal N}}
\newcommand{\EQ}[1]{\begin{equation} #1 \end{equation}}
\newcommand{\AL}[1]{\begin{subequations}\begin{align} #1
\end{align}\end{subequations}}
\newcommand{\SP}[1]{\begin{equation}\begin{split} #1 \end{split}\end{equation}}

\begin{document}

\begin{titlepage}

\setcounter{page}{0}
\renewcommand{\thefootnote}{\fnsymbol{footnote}}

\begin{flushright}
hep-th/0405051\\
SWAT-nnn\\
RIKEN-TH-23
\end{flushright}

\vspace{5mm}
\begin{center}
{\Large\bf
Vacua of $\N{=}1$ Supersymmetric QCD from Spin Chains and 
Matrix Models}

\vspace{8mm}
{
Timothy J. Hollowood${}^1$\footnote{e-mail:
{\tt t.hollowood@swansea.ac.uk}} and Kazutoshi Ohta${}^{1,2}$\footnote{e-mail:
{\tt k.ohta@swansea.ac.uk, k-ohta@riken.jp}}}\\
\vspace{5mm}
${}^1$ {\em Department of Physics\\
University of Wales Swansea\\
Swansea, SA2 8PP, UK}\\

\vspace{5mm}

${}^2$ {\em Theoretical Physics Laboratory\\
The Institute of Physical and Chemical Research (RIKEN)\\
2-1 Hirosawa, Wako\\
Saitama 351-0198, JAPAN}\\

\end{center}

\vspace{8mm}

\centerline{{\bf{Abstract}}}
We consider the vacuum structure of the finite $\N=2$ theory with
$N_f=2N$ fundamental hypermultiplets broken to $\N=1$ by a
superpotential for the adjoint chiral multiplet. We do this in two
ways: firstly, by compactification to three dimensions, in which case
the effective superpotential is the Hamiltonian of an integrable spin
chain. In the second approach, we consider the Dijkgraaf-Vafa 
holomorphic matrix
model. We prove that the two approaches agree as long as the
couplings of the two theories are related in a particular way
involving an infinite series of instanton terms. The
case of gauge group $SU(2)$ with $N_f=4$ is considered in greater detail.

\end{titlepage}
\newpage

\renewcommand{\thefootnote}{\arabic{footnote}}
\setcounter{footnote}{0}

\section{Introduction}

In this work, we examine the breaking of the $\N=2$ $SU(N)$ theory with
$N_f=2N$ massive fundamental hypermultiplets to $\N=1$ by adding a
general tree-level superpotential for the adjoint field of the form
\EQ{
\text{Tr}\,W(\Phi)=\sum_{j=2}^{n+1} \mu_j\text{Tr}\,\Phi^j\ .
\label{tls}
}
Since we describe the situation with general hypermultiplet masses,
the same problem with $N_f<2N$ can be described by taking limits of large
masses where the associated hypermultiplets decouple.

There are several known ways to tackle the problem: factorizing the
Seiberg-Witten curve \cite{Seiberg:1994aj}; using MQCD or deformation of the
Seiberg-Witten curve \cite{deBoer:1997ap,deBoer:2004he}; 
and using matrix models and the 
glueball superpotential (for a review see \cite{Argurio:2003ym} and references
therein). As well
as describing the matrix model technique in some detail, we will
add another approach to the list based on integrable systems. 

The integrable system approach begins with the observation that the
Seiberg-Witten curve of a class of $\N=2$ theories is the spectral
curve of an integrable system. If one compactifies the theory on a
circle to three dimensions and breaks to $\N=1$ as in \eqref{tls}, then
one can write an effective superpotential for a set of scalar fields
that includes the scalar from four dimensions as well as the dual
photons and Wilson lines of the gauge field. The Coulomb branch in
three dimensions is identified with the (complexified) phase space of
the integrable system and the effective superpotential is one of the
Hamiltonians of the integrable system. Hence, vacua correspond to
equilibria of the integrable system \cite{Hollowood:2003ds}. 
The cases which have been
considered in the literature are:

(i) The pure $\N=2$ theory in which case the integrable system is the
affine Toda system \cite{Boels:2003fh,Boels:2003at,Alishahiha:2003hj}.

(ii) The $\N=2^*$ theory (massive adjoint hypermultiplet) in which
case the integrable system is the elliptic Calogero-Moser system 
\cite{Dorey:1999sj,Hollowood:2003ds}.

(iii) The $\N=2$ finite quiver theories in which case the integrable
system is the spin elliptic Calogero-Moser system \cite{Dorey:2001qj}.

(iv) The compactified five-dimensional $\N=2^*$ theory in which case
the integrable system is the elliptic Ruijsenaars-Schneider system
\cite{Hollowood:2003gr}.

(v) The Leigh-Strassler deformed $\N=4$ theory in which case the
integrable system is also the elliptic Ruijsenaars-Schneider system
\cite{Dorey:2002pq,Hollowood:2002nv} 
(in a sense this example is somewhat different because there is no
$\N=2$ parent theory).

In fact the integrable system approach can be adopted in an abstract
sense by considering the problem at the level of action/angle variables
\cite{Hollowood:2003ds} 
even when no concrete realization of the integrable is known
or can easily be written down. 
Now we turn to the theory with fundamental hypermultiplets. The $\N=2$
theory with vanishing beta function, $N_f=2N$, is known to be
associated to a spin-chain integrable system 
\cite{Gorsky:1996hs,Gorsky:1997mw,Gorsky:1997jq,Gorsky:2000px} 
and this suggests that we
can associate the vacua after breaking to $\N=1$ to equilibria of this
integrable system. In Section \ref{Spin Chains} we consider the relation of the
$N_f=2N$ theory with the spin chain integrable system in more detail
and use it to deduce the vacuum structure of the $SU(2)$ theory with
four massive hypermultiplets. 

We note that the formalism can in
principle be extended beyond $SU(2)$ but it becomes harder to find all
the equibilibria. 
In Section \ref{Matrix model}, we use the alternative matrix model
technique to find the glueball superpotentials of these theories, an
approach that readily extends to $N_f<2N$. 
We note that the coupling of the matrix model approach is not equal to
the coupling of the Seiberg-Witten curve. The two are related by a
series of instanton terms:
\EQ{
\tau_1=\tau_2+\sum_{k=0}^\infty c_ke^{\pi ik\tau_2}\ .
\label{relc}
}
The actual relation will be determined in Section 5. In fact it is
known that the coupling of the Seiberg-Witten curve is itself not
equal to the bare coupling of the theory \cite{Dorey:1996bn}, 
rather they are also related
by an expansion of the form \eqref{relc}. 

\section{Semi-Classical Analysis of SQCD Vacua}
\label{Semi-Classical Analysis}

The tree level superpotential of $\N{=}1$ $SU(N)$ supersymmetric QCD
with $N_f$ flavors is obtained from the $\N{=}2$ superpotential by the
addition of a supersymmetry breaking potential for the adjoint scalar field:
\be
W_{\rm tree}(\Phi, Q, \tilde{Q})
=\Tr\tilde{Q}\Phi Q - \Tr M\tilde{Q}Q
+\Tr\,W(\Phi)\ ,
\label{tree QCD potential}
\ee
where $Q$ and $\tilde Q$ are $N\times N_f$ and $N_f\times N$ matrices,
respectively, and $M=\text{diag}(m_1,\ldots,m_{N_f})$ is the mass
matrix. In what follows it will be convenient to formulate the $SU(N)$
theory in terms of the $U(N)$ theory with an explicit Lagrange
multiplier constraint in order to enforce the tracelessness of
$\Phi$. To this end we take the potential for the adjoint to have the form 
\EQ{
W(x)=\sum_{i=1}^{n+1}\mu_ix^i-\xi x\ ,
\label{potential}
}
where $\xi$ is the Lagrange multiplier.

To investigate the vacuum structure one imposes D- and F-flatness
modulo gauge transformations. The latter conditions read
\EQ{
Q\tilde Q+W'(\Phi)=0\ ,\quad
\Phi Q=QM\ ,\quad
\tilde{Q}\Phi=M\tilde Q\ .
\label{ft}
}
As usual solving the D- and F-flatness conditions modulo gauge
transformations is equivalent to only imposing
F-flatness modulo complex gauge transformations. The latter can be used to
diagonalize $\Phi$:
\EQ{
\Phi=\diag(\phi_1,\phi_2,\ldots,\phi_{N})\ .
\label{pdi}
}
This leaves the abelian $U(1)^N$ subgroup as well
as the elements of the
Weyl group which act as permutations of the elements \eqref{pdi}
unfixed.
For generic masses the solutions are as follows.
First of all, for $r=0,\ldots,\text{min}(N,N_f)$, 
\EQ{
\Phi=\diag\big(
m_{I_1},\ldots,m_{I_r},\phi_{r+1},\ldots,\phi_N\big)\ ,
}
where the $I_j$ are distinct elements of the set
$\{1,\ldots,N_f\}$. The $\phi_j$, $j=r+1,\ldots,N$ satisfy
\EQ{
W'(\phi_j)=0\ ,\qquad j=r+1,\ldots,N\ .
}
\EQ{
Q_{jI}=d_j\delta_{II_j}\ ,\quad\tilde Q_{Ij}=\tilde d_j\delta_{II_j}\ ,
\quad j=1,\ldots,r\ ,
}
and 0 otherwise, and where
\EQ{
\tilde{d}_{j}d_{j}+W'(m_{I_j})=0\ ,
 \qquad j=1,2,\ldots,r\ .
}
The fixed gauge symmetry and remaining complex gauge symmetry
can be used to set $\tilde d_i=1$ and so the
solution is at least generically discrete.

The solution breaks the gauge symmetry to 
$\prod_{i=1}^{k}U(N_i)$ with $\sum_{i=1}^{k}N_i=N-r$.\footnote{Note
  that the overall $U(1)$ is frozen out and non-dynamical.} Here, $N_i$ is
the number $\phi_a$ which lie at the $i^\text{th}$ solution of the
polynomial $W'(x)=0$. If $W^*_\text{tree}$ 
is the effective tree-level superpotential in the
vacuum then variable $\xi$ is then determined by setting $\partial
W^*_\text{tree}/\partial\xi=0$. 

Now consider the situation with the simplest potential
$W(x)=\mu x^2/2-\xi x$. In this case, 
\EQ{
\phi_j=\xi/\mu\ ,\qquad
j=r+1,\ldots,N\ .
}
The effective superpotential is
\EQ{
W^*_\text{tree}=\sum_{j=1}^r\Big(\frac\mu2m_{I_j}^2-\xi
m_{I_j}\Big)-\frac{N-r}{2\mu}\xi^2\ .
}
Hence, 
\EQ{
\xi=-\frac\mu{N-r}\sum_{j=1}^rm_{I_j}\ ,
}
and so in the $SU(N)$ theory we have
\EQ{
W^*_\text{tree}=\frac\mu2\sum_{j=1}^rm_{I_j}^2+\frac\mu{2(N-r)}
\Big(\sum_{j=1}^rm_{I_j}\Big)^2\ .
} 
In this vacuum $SU(N)$ is broken to $SU(N-r)$ and for 
$r<N$ the vacuum is characterized by the meson VEV
\EQ{
M= \tilde{Q}Q=
\diag\big(d_1, \ldots,d_r,0,\ldots,0)\ .
}
In the low energy
limit, the theory is an $\N=1$ theory with $SU(N-r)$ gauge group which
conventional reasoning says will confine and there will be $N-r$ vacua. 
Therefore we expect the total number of these vacua is
\EQ{
\sum_{r=0}^{\min(N,N_f)-1}(N-r)\combi{N_f}{r}\ ,
\label{nvac}
}
where $\combi{N_f}{r}$ represents the number of subsets
$\{I_1,\ldots,I_r\}\subset\{1,\ldots,N_f\}$.

The case with $r=N$, which means $N_f\geq N$, is somewhat special but
can only be attained if the $N$-dimensional subset of the 
masses satisfy the constraint $\sum_{i=1}^Nm_{I_j}=0$. In
this case the vacuum is characterizes by a non-zero baryon VEV:
\SP{
B^{I_1\cdots I_{N}}&=
\frac{1}{N!}\epsilon^{j_1\cdots j_{N}}
Q_{j_1I_1}\cdots Q_{j_{N}I_N}=d_1d_2\cdots d_N\ ,\\
\tilde{B}_{I_1 \cdots I_{N}}&=
\frac{1}{N!}\epsilon_{j_1\cdots j_{N}}
\tilde Q_{I_1j_1}\cdots \tilde Q_{I_Nj_N}=\tilde d_1\tilde
d_2\cdots\tilde d_N\ .
}
In these ``Baryonic'' vacua the gauge group is completely broken. 

\subsection{Example: $SU(2)$ with $N_f=4$}

In this section, we consider the finite theory with gauge group $SU(2)$
and $N_f=4$ with a quadratic superpotential $W(x)=\mu x^2/2$, {\it
  i.e.\/}~in our approach we take the $U(2)$ theory with $W(x)=\mu
x^2/2-\xi x$. 

We start with the vacua with $r=0$. In this case there is no Higgsing
and 
\be
\Phi=Q=\tilde Q=0\ .
\ee
At low energy, the theory is described by a pure $\N=1$ theory with
$SU(2)$ gauge group. This will confine and lead to 2 vacua. 
The superpotential vanishes in the classical approximation but we
expect it to receive corrections from fractional instantons with the
characteristic behaviour $e^{\pi i\tau}$.

There are four vacua with $r=1$ with, for $I=1,\ldots,4$,
\be
\Phi=\diag\big(m_I,\xi/\mu)\ ,\quad\xi=-\mu m_I
\label{r=1}
\ee
For each
choice $I$, $Q_{1I}$ and $\tilde Q_{I1}$ are non-vanishing. The $SU(2)$
gauge group is completely broken by the Higgs mechanism and the
classical value of the superpotential is
\EQ{
W_{\rm tree}^*=\mu m_I^2\ .
}

The baryonic vacua can only occur when the masses are non-generic, in
fact one needs $m_I+m_J=0$, for some $I$ and $J$. In this, one can
check that there is a moduli space of vacua.

\subsection{A brane configuration}
\label{brane config}

The vacuum structure that we have established on the basis of a
classical analysis can be readily 
interpreted in terms of the brane configurations.
The $SU(N)$ gauge theory with $N_f$ flavors that we are
considering is realized by a configuration of 
$N$ finite D4-branes stretching between two
NS5-branes and $N_f$ semi-infinite D4-branes emerging from one of the
NS5-branes in Type IIA string theory \cite{Witten:1997sc} (see the review \cite{Giveon:1998sr} and references therein). The world-volumes of D4-branes
and NS5-brane are
along the 01236 and 012345 directions,
respectively, and the gauge theory arises as the effective theory on the 
common directions 0123.

For a theory with $\N{=}2$ supersymmetry, the two NS5-branes are
parallel and the D4-branes corresponding to gauge and flavor symmetry are
freely moving on 45-plane. The $x^4+ix^5$ positions of finite D4-branes
correspond to the (classical) VEV of the adjoint scalar field $\Phi$
and ones of semi-infinite D4-branes correspond to the 
masses of the matters in fundamental representation.

The supersymmetry breaking from $\N{=}2$ to $\N{=}1$ is achieved by a
modification of the relative ``shapes'' of NS5-branes. 
For example, turning on the simple
mass perturbation $W(x)=\mu x^2/2-\xi x$,\footnote{Once again we
introduce the Lagrange multiplier $\xi$ to freeze out the $U(1)$
degree-of-freedom. The variable $\xi$ is fixed by demanding
  the centre-of-mass of the finite D4-branes is zero in the $x$
  plane.} is achieved by rotating one of the 
NS5-branes into the 78 direction \cite{Barbon:1997zu}. Denoting $w=x^7+ix^8$, the 
the rotated NS5-brane is described by the complex line
$w=\mu x-\xi$. The other NS5-brane remains along $w=0$.
Generally, deforming the  $\N{=}2$ theory by turning on the $k+1$-th order
potential as in \eqref{potential} is achieved by demanding that the
shape of one of the NS5-branes is described by
$w=W'(x)$.

The relation between the classification of vacua as in Section
\ref{Semi-Classical Analysis} and
the brane picture is rather direct. A vacuum with given value of $r$,
corresponds to a situation in which $r$ of the semi-infinite D4-branes
match up in $x$-space at $x=m_{I_i}$ $i=1,\ldots,r$, with $r$ of the 
finite D4-branes to make semi-infinite D4-branes which now end on
the left-hand NS5-brane. The remaining $N-r$ finite D4-branes have
to lie at zeros of $W'(x)$. 

For example, for gauge group $SU(2)$
with $N_f=4$, and the quadratic deformation $W(x)=\mu x^2/2-\xi x$, 
there are generically six vacua. There are 2 confining vacua where 
the two finite D4-branes lie at $x=w=0$ giving rise to an unbroken
$SU(2)$ in the IR and the usual count of 2 vacua. There are 4 Higgs vacua 
where one of the semi-infinite D4-branes links up with one of
finite D4-branes at $x=m_I$, $I\in\{1,2,3,4\}$. The remaining
finite D4-brane sits at $x=m_I$. The confining and Higgs vacua
are illustrated in Fig.~1 (a) and (b).  Non-generically there are other
possibilities. For example, if $m_I+m_J=0$ then both finite
D4-branes can link up with 2 of the semi-infinite D4-branes. In this
case, as illustrated in Fig.~1 (c), the brane configuration splits
into two separate pieces.

\begin{figure}
\begin{center}
\includegraphics[scale=0.8]{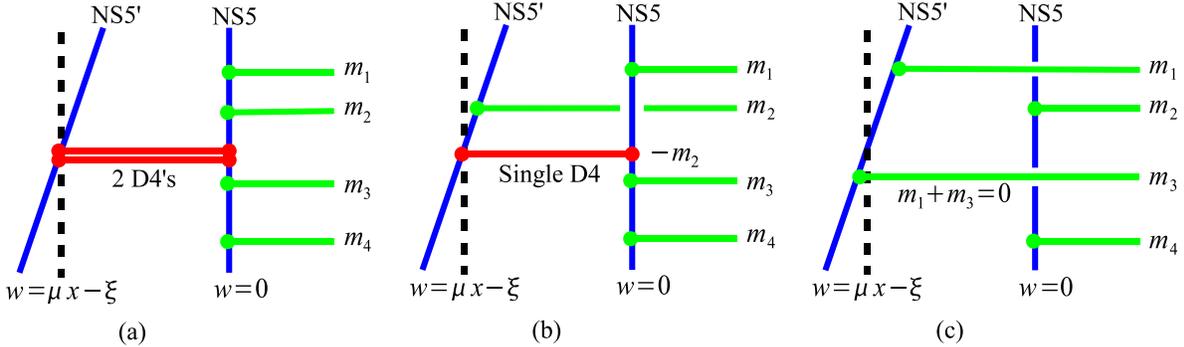}
\end{center}
\caption{The classical brane configuration for
  $(N,N_f)=(2,4)$. In general, the configuration has two finite
  D4-branes between the rotated NS5'-brane and NS5-brane, and four
  semi-infinite D4-branes from the right NS5-brane.
 The bullets on the line denote the end point of
  D4-branes on the NS5-brane. (a) Confining vacua. The number
  of vacua is 2. (b) Higgs vacua. There are 4 different
  choices in total. (c) ``Baryonic'' vacua for non-generic
  masses. The finite D4-branes disappear and brane configuration splits.} 
\end{figure}

\section{Integrable Systems and Vacuum Structure}

\subsection{The general story}

In this section we set out a general picture of the relation between
$\N=2$ theory broken to $\N=1$ and integrable systems. In this story
it is very useful to have in mind the brane configurations of Type IIA
string theory and brane rotation/deformation.

It is well-established that $\N=2$ theories are related to integrable
systems \cite{Gorsky:2000px}.
The method that we describe in this section exploits this
fact. The Seiberg-Witten curve of the theory is identified with the
spectral curve of the integrable system. The moduli space, or Coulomb
branch, of the theory, is therefore identified with the space of
conserved quantities in the integrable system. The Coulomb branch has a
natural set of coordinates which are defined by integrating the
Seiberg-Witten holomorphic 1-form $\lambda$ around a set of canonical
$A$-cycles on the curve $\Sigma$:
\EQ{
a_i=\oint_{A_i}\lambda\ .
\label{defai}
}
Here, $i=1,\ldots,g$, the genus of $\Sigma$. The cycles are chosen so
that the $a_i$ determine the masses of electrically changed BPS states
int he theory. Magnetically charge BPS states involve the dual
quantities
\EQ{
a_i^D=\oint_{B_i}\lambda\ ,
}
where the cycles $B_i$ are chosen to 
have intersections $A_i\cap B_j=\delta_{ij}$. Remarkably, the $a_i$
are precisely the action variables of the integrable system. The
conjugate angle variables do not appear directly in the theory,
however, they are identified with the coordinates $\psi_i$ of a point in the
Jacobian of $\Sigma$. The Jacobian torus is defined as follows (for
a reference on Riemann surfaces see \cite{FK}). 
Let $\omega_j$ be the associated
set of $g$ holomorphic 1-forms (abelian differentials of the 1st
kind) normalized so that
$\oint_{A_j}\omega_k=\delta_{jk}$. The period matrix of $\Sigma_{\rm
  int}$ is the $g\times g$ matrix with elements
\EQ{
\tau_{jk}=\oint_{B_j}\omega_k\ .
}
The Jacobian torus consists of points $\psi_j\in{\bf C}^{g}$ with
the identifications
\EQ{
\psi_j\thicksim \psi_j+n_j+\tau_{jk}m_k\ ,\qquad n_j,m_k\in{\bf Z}\ .
}
As defined the $a_i$ and $\psi_i$ are canonically conjugate:
\EQ{
\{a_i,\psi_j\}=\delta_{ij}\ .
}
In terms of the action angle variables, the dynamics with respect to
any choice of Hamiltonian function $H(a_i)$ is particularly
simple:
\EQ{
\dot a_i=0\ ,\qquad \dot\psi_i=\frac{\partial H}{\partial a_i}\ .
\label{aaf}
}

We can be more explicit by using the fact that the spectral curve of
the integrable system is described by a relation of the form
$F(t,x)=0$ which is nothing but the usual description of the
Seiberg-Witten curve of the field theory coming from the Type IIA
brane configurations described in Section \ref{brane config}. 
In particular, the Seiberg-Witten 1-form
is
\EQ{
\lambda=x\frac{dt}t\ .
}
An important r\^ole is played by the points $P_i$, $i=1,\ldots,k+1$,
where $x=\infty$ at which $\Sigma$ is non-compact. In the brane
configuration these are the points where one of the $k+1$ NS5-branes
goes to infinity.
In order to see
their significance in the integrable system, 
let us consider the flows generated by some
Hamiltonian at the level of the angles variables \eqref{aaf}. It turns out
that each flow can be written in terms of a meromorphic 1-form $\Omega$ on
$\Sigma$ (actually an abelian differential of the 3rd kind)  
which is normalized by the condition\footnote{This can always be
  arranged by adding a suitable linear combination of the $\omega_i$.}
\EQ{
\oint_{A_j}\Omega=0\ .
\label{norm}
}
The angular velocities are then given by integrals around the dual 
cycles:
\EQ{
\dot\psi_i=\oint_{B_i}\Omega\equiv\frac{\partial H}{\partial a_i}\
}
The 1-form $\Omega$ has poles at the points $P_i$ and the asymptotic
expansions at these points are determined by the choice of Hamiltonian
$H$:
\EQ{
\lim_{p\to P_i}\Omega(p)=d\big(U_i(x(p))+{\cal O}(1/x(p))\big)\ ,
\label{oas}
}
where the $U_i(x)$ are polynomials which we will relate to the
tree-level superpotentials which break $\N=2\to\N=1$. In fact in the
brane picture $w=U_i(x)$ describes the bending of the $i^\text{th}$
NS5-brane in the 89 direction asymptotically at infinity. 
The $\N=1$ deformation of the $i^\text{th}$
gauge group factor is associated to a potential
\EQ{
W'_i(x)=U_{i+1}(x)-U_i(x)\ .
\label{viu}
}
Note that a shift in the $U_i(x)\to U_i(x)+f(x)$ is physically
irrelevant. In the integrable system, the polynomials $U_i(x)$
determine specify $\Omega$ uniquely, up to exact terms which
correspond to the freedom to shift each $U_i(x)$ by the same
function. In this sense, only the differences of the $U_i(x)$
matter---as \eqref{viu} makes clear. 

We now relate $\Omega$, modulo exact terms, 
to $H$. The first move is to employ 
the Riemann's bilinear relation to the abelian
differentials\footnote{Note that it will be important below that 
the $a_i$ derivative is taken at fixed $x$.}
\EQ{
\pi_i\equiv\frac{\partial\lambda}{\partial
  a_i}
\label{defpi}
}
and $\Omega$:
\EQ{
\frac1{2\pi i}
\sum_{i=1}^{g}
\left(
\oint_{A_i}\Omega\,\oint_{B_i}\pi_j-\oint_{B_i}\Omega
\,\oint_{A_i}\pi_j
\right)
=\sum_{i=1}^{k}{\rm Res}_{P_i}(U_i(x)\pi_j)\ ,
}
Now we use the normalization condition \eqref{norm} and the fact that
\eqref{defai} and \eqref{defpi} imply that
\EQ{
\oint_{A_i}\pi_j=\delta_{ij}\ ,
}
to deduce
\EQ{
-\oint_{B_i}\Omega
=\sum_{i=1}^{k+1}{\rm Res}_{P_i}(U_i(x)\pi_i)
=\frac{\partial}{\partial a_i}
\sum_{i=1}^{k+1}{\rm Res}_{P_i}(U_i(x)\log t\,dx)\ .
}
In the last line we used the fact that the $a_i$ derivative is at
constant $x$ and so could be pulled outside the integral.
Hence, by comparing this with \eqref{aaf} we deduce that
\EQ{
H=-\sum_{i=1}^{k}{\rm Res}_{P_i}(U_i(x)\log t\,dx)
\label{hamd}
}

At this point we can
investigate the equilibrium configurations of the integrable system
which ought, by our general reasoning, to be associated to vacua of
the QFT. For an equilibrium,
\EQ{
\frac{\partial H}{\partial a_i}=0\ \implies\ \oint_{B_i}\Omega=0\ .
}
Given the normalization condition this implies that for an equilibrium
$\Omega=dw$, for some meromorphic function $w$ with singularities at
$P_i$ with asymptotic behaviour from \eqref{oas}, $w\to
U_i(x)+\cdots$. Hence the existence of a vacuum of the gauge theory can be
phrased as the existence of a certain meromorphic function on the
curve of the theory. The relation to the brane configuration should
now be obvious. The points $P_i$, as we have said, 
represent the asymptotic positions of
the NS5-branes. The meromorphic function $w$ describes the
deformation of the Seiberg-Witten curve into the 89-plane
associated with the breaking of $\N=2$ to $\N=1$. The superpotential
\eqref{hamd} at the critical point can be written as an integral over
the Riemann surface
\EQ{
H^\star=-\frac1{2\pi i}\int_\Sigma \log t\,dw\wedge dx\ ,
}
from which one can relate to Witten's MQCD superpotential
\cite{Witten:1997ep} (see
\cite{deBoer:2004he} for a discussion of the relation).

For our particular example, there are two NS5-branes and hence two
points at infinity $P_1\equiv\infty^+$ and $P_2=\infty^-$. We can
choose $U_2=0$ and $U_1(x)=W'(x)$. By integrating-by-parts in
\eqref{hamd}, we have a Hamiltonian and hence an effective superpotential
\EQ{
W_\text{int}=\text{Res}_{\infty^+}\Big(W(x)\frac{dt}t\Big)\ ,
\label{esph}
}

Notice that we have developed the theory of the integrable system
entirely in the language of action/angle variables. However, for
certain examples we can write down the Hamiltonians in terms of other
kinds of variables; for instance, the positions and momenta of
particles, or spin variables and mixtures thereof. We do this for 
the theory with $N_f=2N$ fundamental hypermultiplets in the next section.

\subsection{Spin Chains}
\label{Spin Chains}

In this section, we describe how to extend the integrable system
approach to describe the $\N=2$ theories with fundamental matter. The
integrable system that is associated to the finite theory, with
$N_f=2N$, has been identified in 
\cite{Gorsky:1996hs,Gorsky:1997mw,Gorsky:1997jq,Gorsky:2000px} as a
certain twisted $SU(2)$ spin chain. This system describes the
interactions of a set $N$ spins 
$\vec S^{(i)}$, $i=1,\ldots,N$, which are 3-vectors
\EQ{
\vec S^{(i)}=\big( S^{(i)}_1,S^{(i)}_2,S^{(i)}_3\big)\ .  
\label{spin}
}
The overall length of each vector is non-dynamical:
\EQ{
\vec S^{(i)}\cdot\vec S^{(i)}=k_i^2\ .
\label{cas}
}
The parameters $k_i$ will be associated to masses in due course. 

The integrable system has a $2\times2$ 
``Lax matrix'' defined at each ``site'':
\EQ{
L_i(x)=\begin{pmatrix} x+\lambda_i+S^{(i)}_3 & S^{(i)}_1-iS^{(i)}_2\\
S^{(i)}_1+iS^{(i)}_2 & x+\lambda_i-S^{(i)}_3\end{pmatrix}
}
The ``transfer matrix'' is then defined to be
\EQ{
T(x)=L_1(x)\cdots L_N(x)V
}
and we have allowed for a ``twisting'' described by the constant
matrix $V$. Unlike
\cite{Gorsky:1996hs,Gorsky:1997mw,Gorsky:1997jq,Gorsky:2000px}, 
we will make the choice
\EQ{
V=\begin{pmatrix}-h&0\\ 0& h+2\end{pmatrix}\ ,
}
where $h$ is a parameter that we will relate to the UV coupling of the theory.
The integrable system has a {\it spectral curve\/} $\Sigma$ defined by the
algebraic relation between $x$ and another variable $t$:
\EQ{
F(x,t)=\det\big(T(x)-t{\bf 1}\big)=0\ .
}
This curve is easily seen to have genus $N-1$.
The relation between the spectral curve and the integrable system is
well known. The moduli space of the curve parameterizes the space of
conserved quantities of the system. There are $3N$ variables subject
to $N$ constraints \eqref{cas}; hence, the dimension of the phase space
is $2N$. In particular, the action variables
are given by integrating the 1-form $xd\log t$ around a set of basis
1-cycles on the curve. The conjugate angle variables are associated to
a point in the Jacobian of $\Sigma$.\footnote{Note that the curve
  actually has genus $N-1$ and so there is an apparent mismatch between the
  number of action-angle variables, $N$ each, and the number of
  basis 1-cycles, or dimension of the Jacobian, of the curve,
  $N-1$. We will resolve this conflict below.}

The curve $\Sigma$ has the form
\EQ{
t^2-2tP(x)-h(h+2)Q(x)=0\ ,
\label{swc}
}
where
\EQ{
Q(x)=\prod_{I=1}^{2N}(x-m_I)
\label{masq}
}
with
\EQ{
m_{2i-1}=-\lambda_i-k_i\ ,\qquad m_{2i}=-\lambda_i+k_i\ ,
\label{mass}
}
$i=1,\ldots,N$, and where
\EQ{
P(x)=x^N+H_1x^{N-1}+H_2x^{N-2}+\cdots+H_N\ .
\label{pdef}
}
Notice that the form of the curve is identical to the Seiberg-Witten
curve of the $\N=2$ theory with $SU(N)$ gauge group and $N_f=2N$
fundamental hypermultiplets, and where $h$ is expressed in terms of the
coupling by \cite{Argyres:1995wt,Argyres:1996eh}
\EQ{
h(\tau)=\frac{2\lambda(\tau)}{1-2\lambda(\tau)}\ ,
\label{swh}
}
where $\lambda(\tau)$ involves the automorphic function 
\EQ{
\lambda(\tau)=16q\prod_{n=1}^\infty\left(\frac{1+q^{2n}}
{1+q^{2n-1}}\right)^8\ ,
}
and $q=e^{\pi i\tau}$.
In particular, at weak coupling $h=32q+\cdots$.
The parameters $m_I$ in \eqref{masq} are not directly the bare masses, in
fact
\EQ{
m_I=m_I^b+h\sum_{I=1}^{2N}m_I^b/(2N)\ .
}
In addition, the coupling $\tau$ in the curve is not equal to the bare
coupling of the theory \cite{Dorey:1996bn}. In general they are related by an
instanton expansion of the form \eqref{relc}.

In \eqref{pdef} the $H_i(\vec S^{(j)})$ are a basis for the 
Hamiltonians of the integrable system. Explicitly, the first two
Hamiltonians are 
\EQ{
H_1=\sum_i\big(\lambda_i-(1+h)S^{(i)}_3\big)
}
and
\EQ{
H_2=\sum_{i<j}\Big(\lambda_i\lambda_j+i(1+h)\big(S^{(i)}_2S^{(j)}_1-
S^{(i)}_1S^{(j)}_2\big)+\vec S^{(i)}\cdot\vec
S^{(j)}-(1+h)\big(\lambda_iS^{(j)}_3+\lambda_jS^{(i)}\big)\Big)\ .
}

\subsection{Breaking to $\N=1$}

Now that we have established the precise relation between the twisted 
spin chain and the $\N=2$ field theory, we now consider the problem of
breaking to $\N=1$ using the integrable system. The procedure is well
established \cite{Dorey:1999sj,Dorey:2001qj}. 
First of all, consider compactifying the theory 
to three dimensions on a circle of finite radius $R$. In the
3-dimensional effective theory, the gauge field in the $U(1)^{N-1}$
unbroken gauge group can be ``dualized'' to scalar fields. These
fields, along with the Wilson lines of the $U(1)^{N-1}$ around the
circle amass into complex scalar field. The $N-1$ complex scalar
fields are then naturally valued on a complex torus; in fact,
precisely the Jacobian torus of $\Sigma$. In summary the Coulomb
branch of the theory in three dimensions is identified with the
complexified phase space of the integrable system and the split
between the action-angle variables describes the moduli of the Coulomb
branch of the four-dimensional theory plus the dual photons and Wilson
lines. 

Breaking to $\N=1$, as in \eqref{tls}, is described by an effective
superpotential on the Coulomb branch of the 3-dimensional theory. The
resulting superpotential is simply, as described in the last section in
detail, one of the Hamiltonians of the
integrable system. From \eqref{esph}, we have the following expression
for the effective superpotential 
\EQ{
W_\text{int}=\text{Res}_{\infty^+}W(x)\frac{dt}t=
\text{Res}_{\infty^+}W(x)\left(
\frac{P'}y+
\frac{h(h+2)Q'}{2y(y+P)}
\right)\ .
\label{intsp}
}
In particular, since $W(x)$ contains the term $-\xi x$ we have the
constraint 
\EQ{
\text{Res}_{\infty^+}x\left(
\frac{P'}y+
\frac{h(h+2)Q'}{2y(y+P)}
\right)
=0\ .
}
This can be written
\EQ{
H_1(\vec S^{(j)})=\frac h2\sum_{I=1}^{2N}m_I=-(\lambda_1+\lambda)h\ .
\label{ham1z}
}
At the level of the
integrable system experience with the finite
$U(N)^k$ quiver theories \cite{Dorey:2001qj}, 
where the relative $U(1)$ factor are
also IR free, suggests that the correct way to freeze out the $U(1)$
factor is via a Hamiltonian reduction. So as well as imposing
\eqref{ham1z} on the phase space, one also takes  a quotient by the
conjugate variable to $H_1$. Notice that $H_1$ generates a simultaneous
rotation of the spins $\vec S^{(i)}$ around the $S^{(i)}_3$ axes and
so the conjugate variable is this angle. 
In practice it is easier not to perform the
quotient explicitly and as a consequence we shall find a degeneracy in
our critical points corresponding to this unwanted degree-of-freedom.

\subsection{Example: $SU(2)$}

In this section, we consider the vacuum structure for the case of
gauge group $SU(2)$ and the quadratic mass deformation $W(x)=\mu
x^2/2-\xi x$. After imposing the constraint, and using \eqref{intsp}, 
we find an effective superpotential
\EQ{
W_\text{int}=-\frac{\mu(H_2+h\sum_{I<J}m_Im_J/2)}{2(h+1)}\ .
}
So vacua are given by critical point of $H_2$.
First of all, we solve the
constraint \eqref{ham1z} explicitly by choosing
\EQ{
S^{(1)}_3=y\ ,\qquad S^{(2)}_3=\lambda_1+\lambda_2-y\ .
}
and then impose the constraints \eqref{cas} with Lagrange multipliers
$\xi_i$, $i=1,2$. In order the that the four equations that result from the
derivatives of $H_2$ with respect to $S^{(1)}_{1,2}$ and $S^{(2)}_{1,2}$, do
not imply that the four aforementioned variables vanish (which would
not be compatible with the constraints \eqref{cas}), we need
\EQ{
\xi_1\xi_2+\frac{h(h+2)}4=0\ .
} 
{}From this we find 
\EQ{
S^{(1)}_1=\frac{-S^{(2)}_1+i(1+h)S^{(2)}_2}{2\xi_1}\ ,\qquad
S^{(1)}_2=\frac{-i(1+h)S^{(2)}_1-S^{(2)}_2}{2\xi_1}\ .
}
The equation $\partial H_2/\partial y=0$ can then be solved for $y$.
This leaves the two constraints \eqref{cas} which imply that the
Lagrange multiplier $\xi_1$ satisfies a sixth-order polynomial equation
\SP{
&-64k_1^2\xi_1^6+128k_1^2\xi_1^5+16(2(h^2+2h-2)k_1^2+4\lambda_1^2+h(-2k_2^2
+6\lambda_1^2+2\lambda_2^2)+\\ &h^2(-k_2^2+2\lambda_1+2\lambda_2^2))\xi_1^4
+32h(h+2)(-k_1^2+k_2^2(1+h)(\lambda_1^2-\lambda_2^2))\xi_1^3\\
&+4h(h+2)(-h(h+2)k_1^2+2((h^2+2h-2)k_2^2+(h+1)(2\lambda_2^2+
h(\lambda_1^2+\lambda_2^2))))\xi_1^2\\
&-8h^2(h+2)^2k_2^2\xi_1-h^3(h+2)^3k_2^2=0\ .
}
Note that each of the 6 solutions is degenerate due to simultaneous  
rotations in the 
$(S^{(1)}_1,S^{(1)}_2$ and $(S^{(2)}_1,S^{(2)}_2)$ planes. However,
this is the expected and unphysical degeneracy in the conjugate variable
to $H_1$.

So the final result is that there are six vacua, which is the number expected
\eqref{nvac}. For generic values of the parameters it is not
possible to write down explicitly the critical value of the
superpotential. However, let us 
consider their weak coupling, $h\to0$ limit.
There are three pairs of solutions:

(i) $\xi_1=1\pm\lambda_1/k_1+{\cal O}(h)$ and $\xi_2={\cal O}(h)$, which gives
\SP{
\vec S^{(1)}&=\Big(-\frac{\sigma-i\eta}{2(1\pm\lambda_1/k_1)},
\frac{-i\sigma-\eta}{2(1\pm\lambda_1/k_1)},
\mp k_1\Big)+{\cal O}(h)\ ,\\
\vec S^{(2)}&=\Big(\sigma,\eta,\lambda_1+\lambda_2\pm k_1\Big)+{\cal
  O}(h)\ ,
}
where $\sigma^2+\eta^2=k_2^2-(\lambda_1+\lambda_2\pm k_1)^2$.

(ii)  $\xi_1={\cal O}(h)$ and 
$\xi_2=1\pm\lambda_2/k_2+{\cal O}(h)$, which gives
\SP{
\vec S^{(1)}&=\Big(\xi,\eta,\lambda_1+\lambda_2\pm k_2\Big)+{\cal
  O}(h)\ ,\\
\vec
S^{(2)}&=\Big(-\frac{\sigma-i\eta}{2(1\pm\lambda_2/k_2)},\frac{-i\sigma-\eta}
  {2(1\pm\lambda_2/k_2)},\mp k_2\Big)+{\cal O}(h)\ ,
}
where $\xi^2+\eta^2=k_1^2-(\lambda_1+\lambda_2\pm k_2)^2$.

(iii) $\xi_1=\pm((k_2^2-\lambda_2^2)h/2(\lambda_1^2-k_1^2))^{1/2}+
{\cal O}(h)$, which gives
\SP{
\vec S^{(1)}&=\Big(-\frac{\sigma-i\eta}{2\xi_1},
\frac{-i\sigma-\eta}{2\xi_1},
\lambda_1\Big)+{\cal O}(h)\ ,\\
\vec S^{(2)}&=\Big(\sigma,\eta,\lambda_2\Big)+{\cal
  O}(h)\ ,
}
where $\sigma^2+\eta^2=\lambda_2^2$.

In all these solutions the degeneracy due to rotations in $(\xi,\eta)$
corresponds to the angle variable conjugate to $H_1$ and is not
physically relevant.
Clearly the four vacua (i) and (ii) are the Higgs vacua with an
expansion in instantons $h\sim e^{\pi i\tau}$ at weak coupling. For these 
vacua we can write the first two terms in the weak coupling expansion
as
\EQ{
W^*_\text{int}=\mu m_I^2-\frac{\mu\prod_{J=1}^4
(m_I+m_J)}{8m_I^2}h+{\cal
  O}(h^2)
\label{inth}
}
The two vacua (iii) are identified with the confining vacua since they
have an expansion in terms of $h^{1/2}$, in other words, finite
instantons, at weak coupling. In these vacua
\EQ{
W^*_\text{int}=\pm 2\mu \big(m_1m_2m_3m_4h/2\big)^{1/2}+{\cal O}(h)\ .
\label{intc}
}  

These results may be checked very directly by factorizing the
Seiberg-Witten curve. Defining the curve as 
\EQ{
G(x,t)=t-2P(x)-h(h+2)Q(x)/t\ .
}
For the $SU(2)$ case this genus one curve degenerates to genus zero
when
\EQ{
\frac{\partial G}{\partial t}=\frac{\partial G}{\partial x}=0\ .
}
These equations can be reduced to a single sixth-order polynomial 
equation for $x$:
\EQ{
h(h+2)Q'(x)^2+4Q(x)P'(x)^2=0\ ,\qquad t=-h(h+2)\frac{Q'(x)}{2P'(x)}\ .
}
The six solutions correspond to the four Higgs and two confining vacua
and one can verify the result above.

\section{The Holomorphic Matrix Model}
\label{Matrix model}

The effective superpotential of
our supersymmetric field theory can also be calculated from a holomorphic 
matrix model in the way describing originally by Dijkgraaf and Vafa
\cite{Dijkgraaf:2002fc} and then extended to 
include fundamental matter in \cite{Argurio:2002xv}
(see also the review \cite{Argurio:2003ym} and references therein). 

The matrix model is defined by the partition function
\EQ{
Z=\int[d\hat{\Phi}][d\hat{Q}][d\hat{\tilde{Q}}]
 e^{-\frac{1}{g_s}W_{\rm tree}(\hat{\Phi},\hat{Q},\hat{\tilde{Q}})}\ ,
}
where $W_{\rm tree}(\hat{\Phi},\hat{Q},\hat{\tilde{Q}})$
 has the same form as the field
theoretical tree level superpotential, but each $\hat{\Phi}$, $\hat{Q}$ and
$\hat{\tilde{Q}}$ are now $\hat{N} \times \hat{N} $,
 $\hat{N} \times N_f$
 and $N_f\times\hat{N} $
matrices, respectively, not fields. Here $\hat{N} $ is a
number of ``colors'' to be taken the large $\hat{N} $ with fixed
physical quantities of $N$, $N_f$ and $S=g_s \hat{N} $. The integral
is to be understood in a holomorphic sense.

The matrices $\hat Q$ and $\hat{\tilde Q}$ appear quadratically and
can be integrated out. Gauge transformation can then be used to
diagonalize $\Phi$ at the expense of introducing a gauge-fixing
determinant---the Vandermonde determinant. Denoting the eigenvalues of 
$\Phi$ as $\lambda_i$,
$i=1,2,\ldots,\hat{N}$, the partition function becomes
\be
Z=\int\prod_{i=1}^{N}d\lambda_i\,
\frac{\prod_{i\neq j}(\lambda_i-\lambda_j)}
{\prod_{I=1}^{N_f}\prod_{i=1}^{N}(\lambda_i-m_I)}
e^{-\frac{1}{g_s}\sum_{i=1}^NW(\lambda_i)}.
\label{partition function}
\ee

We now take the limit $\hat N\to\infty$ and $g_s\to0$ with $S=g_s\hat
N$ fixed. In this limit,  it is appropriate to introduce a 
density of eigenvalues
\be
\rho(\lambda)=\frac{1}{\hat{N} }\sum_{i=1}^{\hat{N} }
\delta(\lambda-\lambda_i) ,
\ee
which is normalized as
\be
\int d\lambda\, \rho(\lambda)=1\ .
\ee
The free-energy of the model has the expansion
\be
-\ln Z = \frac{1}{g_s^2}{\cal F}_{\chi=2}
+\frac{1}{g_s}{\cal F}_{\chi =1}\ ,
\ee
where
\AL{
{\cal F}_{\chi=2} & =S\int d\lambda\,\rho(\lambda)W(\lambda)
-S^2\int d\lambda d\lambda'\,
\rho(\lambda)\rho(\lambda') \log (\lambda-\lambda')\ ,\label{F2}\\
{\cal F}_{\chi=1}
& =S\sum_{I=1}^{N_f}\int d\lambda\,\rho(\lambda)\log(\lambda-m_I),
\label{F1}
}
are a contribution from the planer diagrams (sphere) and diagram with
one boundary, respectively.

To leading order, the density of eigenvalues is determined from the
saddle-point equation 
\EQ{
2\int d\lambda'\frac{\rho(\lambda')}{\lambda-\lambda'}=\frac1SW'(\lambda)\
.
\label{saddle point eq}
}
In the classical limit ($S\to0$), the eigenvalues sit at one of the
critical points of $W(x)$. In other words, a classical configuration of
eigenvalues can be described by the filling fraction $\hat N_i/\hat N$
of the $i^\text{th}$ critical point. We are assuming that $W(x)$ is a
polynomial of degree $n+1$. However, the saddle-point
equation includes interaction terms coming from the Vandermonde determinant
as well as from integrating out the matter matrices. At this point,
we apply the usual hypothesis which states 
that the effect of these terms is to pull the
eigenvalues away from the critical points along a set of 
open contours $\CC_i$, $i=1\,\ldots,n$, in the neighbourhood of each
critical point. 
The saddle-point equations \eqref{saddle point eq} can be solved by
introducing a {\it resolvent\/}
\be
\omega(x)=\int_{\CC} d\lambda\,\frac{\rho(\lambda)}{x-\lambda}\ ,
\label{resolvent}
\ee
where $\CC=\cup_{i=1}^n\CC_i$. Note that $\omega(x)$ is an analytic function
over the complex plane with cuts along
$\CC_i$. The density can be recovered from $\omega(x)$ as the
discontinuity across the cuts $\CC$:
\EQ{
\omega(x+\epsilon)-\omega(x-\epsilon)=-2\pi i\rho(x)\ ,\quad x\in\CC\ ,
}
where $\epsilon$ is a suitable infinitesimal, so that $x\pm\epsilon$
lie just above and below the cut, respectively.
In terms of $\omega(x)$, the saddle-point equation becomes
\EQ{
\omega(x+\epsilon)+\omega(x-\epsilon)=\frac1SW'(x)\ ,\qquad x\in\CC\ .
\label{spe}
}
Since $\omega(x)$ goes to 0 at infinity, 
the solution to the saddle-point equation \eqref{spe} is immediate:
\EQ{
2S\omega(x)=W'(x)-\sqrt{W'(x)^2+f(x)}\ ,
}
where $f(x)$ is an arbitrary polynomial of degree $n-1$. This latter
requirement ensures
\EQ{
\lim_{x\to\infty}2S\omega(x)={\cal O}(1/x)\ .
}
It is useful to define
\EQ{
y^2=\big(W'(x)-2S\omega(x)\big)^2=W'(x)^2+f(x)\ .
}
The two sheets of $y$ describe a hyper-elliptic Riemann surface
$\Sigma_\text{mm}$ with genus---at least generically---of $n-1$. 

The $k$ constants in $f(x)=\sum_{j=1}^{n}b_jx^{j-1}$ 
are the moduli of the saddle-point
solution, {\it i.e.\/}~of the auxiliary surface $\Sigma_\text{mm}$. 
These $k$ moduli represent the freedom to choose the $k$
filling fractions at each cut and we may change variables
$\{b_i\}\to\{S_i\}$, where
\EQ{
S_i=g_s\hat N_i=\int_{\CC_i}\rho(x)dx=-\frac 1{4\pi i}
\oint_{A_i}y(x)dx\ ,
\label{defsi}
}
where $A_i$ is a contour which encloses the cut $\CC_i$. In
particular,
\EQ{
S=\sum_{i=1}^nS_i=-\sum_{i=1}^n\frac 1{4\pi i}
\oint_{A_i}\,y(x)dx=\frac1{4\pi i}\oint_{\infty^+}y(x)\,dx
=-\frac{b_N}{4\mu_{N+1}}\ ,
}
where we pulled the contour off the cuts to the point at infinity
$\infty^+$ on the top sheet ($y=\sqrt{W'(x)^2+f(x)}$).

We now have all the ingredients to define the glueball
superpotential which is the effective superpotential of the physical
field theory, in a particular classical vacuum,
where the $S_i$ are interpreted as fields:
\be
W_{\rm gb}=\sum_{i=1}^{n}
\left(
N_i\frac{\del {\cal F}_{\chi=2}}{\del S_i}+\pi i \tau S_i
\right)
+{\cal F}_{\chi=1}\ .
\label{gbsp}
\ee
Here, $\tau$ is the bare coupling of the field theory and the
classical vacuum is specified by the
$N_i$, $\sum_{i=1}^nN_i=N$, 
the number of eigenvalues of the adjoint field $\Phi$  lying at the
$i^\text{th}$ critical point of $W(x)$. 

To evaluate the
derivative of ${\cal F}_{\chi=2}$ in \eqref{gbsp}, one can relate it
to the
variation of the free energy in bringing an eigenvalue in
from infinity to the point $e_i$ at the end of the $i^\text{th}$
cut. Since the force on eigenvalue is $y$ we have
\be
\frac{\del {\cal F}_{\chi=2}}{\del S_i}
=\int_{\infty^+}^{e_i}y\,dx\ .
\label{fds}
\ee
The contribution from the matter fields, the third term in
\eqref{gbsp}, is easily evaluated, again one can show \cite{Cachazo:2003yc}
\be
{\cal F}_{\chi=1}=\tfrac12\sum_{I=1}^{N_f}\Big(-\int^{\infty^-}_{m_I}y\,dx
+W(m_I)\Big)\ .
\label{int matter}
\ee
In \eqref{fds} and \eqref{int matter}, the integrals are actually
divergent at infinity and need to be regularized. However, since we
will work ultimately in the finite theory where the divergences cancel
we shall not explicitly describe the details of the regularization process. 
In \eqref{int matter}, the point with $x=m_I$
can be either on the top or bottom sheet of $\Sigma$ and this freedom
to choose is necessary in order to describe all the vacua of
the physical theory \cite{Cachazo:2003yc}. Since we will later choose
all the $m_I$ on the bottom sheet when we engineer the Seiberg-Witten
curve, we take the integral out to the point at infinity on the bottom
sheet. If we need to specify whether $x=m_I$ is on the top or bottom
sheet we write $m^\pm_I$, respectively.

Using \eqref{fds} and \eqref{int matter}, 
the glueball superpotential \eqref{gbsp} takes the form
\EQ{
W_{\rm gb}=-\sum_{i=1}^nN_i \int_{e_i}^{\infty^+} y\,dx-\frac\tau4
\sum_{i=1}^n\oint_{A_i}y\,dx-\tfrac{1}{2}\sum_{I=1}^{N_f}
\left(\int_{m_I}^{\infty^-}
y\,dx-W(m_I)\right)\ .
\label{ep integral}
}
We now specialize to the case $N_f=2N$ and in this case the potential
divergences at the infinite points cancel and the regular can be removed.

\subsection{Engineering the Seiberg-Witten curve}

The Seiberg-Witten curve itself can be extracted from the matrix model
by choosing a suitably generic potential $W(x)$ with order
$N+1$:
\EQ{
W'(x)=\mu_{N+1}\prod_{i=1}^N(x-\zeta_i)\ .
}
This potential acts as a probe for the $\N=1$ vacuum 
which can zero-in on any point on the Coulomb branch of the $\N=2$
theory. 

The matrix model curve $\Sigma_\text{mm}$ for an $N$-cut solution
at the critical point of the glueball superpotential, with one physical
eigenvalue at each critical point $N_i=1$, is then
identified with the Seiberg-Witten curve where the $\{\zeta_i\}$ are
related to the moduli $\{a_i\}$
of the $\N=2$ Coulomb branch as in \eqref{swc}. In addition the points
$x=m_I$ are taken on the bottom sheet, which we indicate as $m^-_I$.
We now proceed to find the
critical point of the glueball superpotential in this case.

It is important for what follows that the $N-1$ quantities 
\EQ{
\frac{\partial}{\partial b_i}y\,dx=\frac{x^{i-1}}{2y}\,dx\ ,
}
for $i=1,\ldots,N-1$, form a basis for the abelian differentials of the first
kind on $\Sigma$. We will choose another basis for the abelian
differentials $\{\omega_i\}$, $i=1\ldots,N-1$, normalized by
\EQ{
\oint_{A_i}\omega_j=\delta_{ij}\ ,\quad i,j=1,\ldots,N-1\ .
}
In addition,
\EQ{
\frac{\partial}{\partial b_i}y\,dx=\frac1{
2\mu_{N+1}}\tau_{\infty^+\infty^-}\ ,
}
where on the right-and side we have 
$\tau_{p_1p_2}$, the meromorphic 1-form with 
simple poles at $p_1$ and $p_2$ with 
residues $\mp1$, respectively. 

The glueball superpotential \eqref{ep integral} (with $N_i=1$ and
$N_f=2N$) can be written
\EQ{
W_\text{gb}=-\frac12\sum_{i=1}^{N}
\int_{\hat B_i}y\,dx-\frac{\pi i\tau}{4\mu_{N+1}}b_N
-\frac{1}{2}\sum_{I=1}^{2N}
\left(\int_{m_I^-}^{\infty^-} y\,dx-W(m_I)\right)\ ,
\label{ep integral2}
}
where $\hat B_i$ is the contour that goes from $\infty^-$ through the
$i^\text{th}$ cut to $\infty^+$. To reach this expression we used the
fact that $\int_{e_i}^{\infty^+}y\,dx=\tfrac12\int_{\hat B_i}y\,dx$.
As we have already explained this expression must be 
regularized, although at the end in the $N_f=2N$ case the regulator
can be removed. The careful treatment of the regularization process is
given in \cite{Cachazo:2003yc}.

\begin{figure}[ht]
\begin{center}
\includegraphics[scale=0.6]{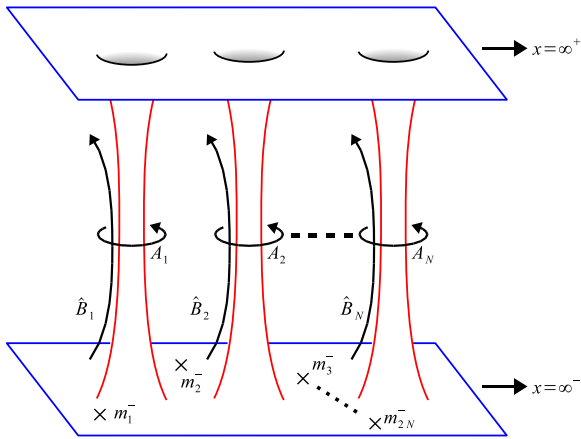}
\end{center}
\caption{The integral paths for the effective superpotential
  calculation. The curve has two sheets 
connecting $N$ cuts
  and there are two
  distinguish infinite points, which we denote $\infty^+$
  and $\infty^-$. The integral cycles are defined in the figure.}
\end{figure}

Taking the derivative of \eqref{ep integral2} with respect to
$b_i$, $i=1,\ldots,N-2$, (we will consider the $b_{N-1}$ equation
below) and taking suitable linear combinations to
change basis between $\{x^idx/y\}$ and $\{\omega_i\}$, we have
\EQ{
0=-\sum_{j=1}^N\int_{\hat B_j}\omega_i
-\sum_{I=1}^{2N}\int_{m_I^-}^{\infty^-}\omega_i
=-N\int_{\infty^-}^{\infty^+}\omega_i
-\sum_{I=1}^{2N}\int_{m_I^-}^{\infty^-}\omega_i\ ,
\label{divisor}
}
where in the last line we have equality up to an element of the
Jacobian lattice, $p_i+\tau_{ij}q_j$, $p_i,q_i\in\Z$. 

According to Abel's theorem, \eqref{divisor} implies that 
there exists a meromorphic function on $\Sigma$ with a divisor
$(\infty^+)^{-N}(\infty^-)^{-N}m_1^-\cdots m_{2N}^-$. This function is
explicitly
\EQ{
t=P(x)+y\ ,
}
where $P(x)=\prod_{i=1}^N(x-a_i)$ 
is a polynomial of degree $N$. In order that $t$ has zeros at
$x=m_I$ on the bottom sheet implies that
\EQ{
P(x)+\gamma Q(x)=W'(x)^2+f(x)\ .
}
where $Q(x)=\prod_{I=1}^{2N}(x-m_I)$.
These are $2N$ equations that determine $\{a_i,b_i\}$. There
is one remaining unknown $\gamma$ which will be fixed below.
Hence, at the critical point the curve can be written
\EQ{
y^2=P(x)^2+\gamma Q(x)\ ,
\label{mmc}
}
We recognize 
this as precisely the Seiberg-Witten curve \eqref{swc} and in particular
the $\{a_i\}$ are the moduli on the Coulomb branch. 

The remaining equation is obtained by taking the derivative with
respect to $b_N$. After using a Riemann bilinear relation with $d\log
t$ and $\partial(y\,dx)/\partial b_{N-1}$ one obtains
\EQ{
\tau=\frac{t(\infty^-)}{t(\infty^+)}
=\frac1{\pi i}\log\frac{1-\sqrt{1+\gamma}}{1+\sqrt{1+\gamma}}\ . 
}
Note that this expression, 
for the case $N_f<2N$, requires
regularization, however, in the finite theory this is not necessary
and a finite answer is obtained. Expressing $\gamma=h(h+2)$, as in
\eqref{swc}, we find
\EQ{
h=-\frac{2e^{\pi i\tau}}{1+e^{\pi i\tau}}\ .
\label{mmh}
}
Comparing with \eqref{swh}, we see that the matrix model coupling,
which we now denote as $\tau_\text{mm}$, cannot be
equal to the Seiberg-Witten coupling $\tau_\text{SW}$ 
directly, rather they are related by an expression of the form \eqref{relc}.

We can now see that the matrix model approach to the vacuum problem
agrees precisely with the integrable system treatment. Firstly the
matrix model curve and Seiberg-Witten curve are identical, \eqref{swc}
and \eqref{mmc}. Secondly one can compare the critical value of the
superpotential. 
Using once again a Riemann bilinear identity for $d\log t$ and $y\,dx$
one finds that the critical value of the glueball superpotential is
\EQ{
W^*_\text{gb}={\rm Res}_{\infty^+}\Big(W(x)\frac{dt}t\Big)
}
a result that agrees precisely with $W^*_\text{int}$ in
\eqref{esph}. 

\subsection{Effective superpotential for the quadratic potential}

In this section, we analyze the effective superpotential for the
quadratic potential $W(x)=\mu x^2/2$. This corresponds to looking at
the one cut solution of the matrix model. 
Actually, in order to compare with the results of Section \ref{Spin Chains}
we have
to take into account the behaviour of the overall $U(1)$ factor. The
matrix model calculation on the face of it involves the gauge group
$U(N)$ rather than $SU(N)$. However, the $U(1)$ factor is, as we have
described earlier, actually frozen out and is non-dynamical. However,
in the matrix model approach the trace of $\Phi$ is non-vanishing. One
finds
\EQ{
\langle\Tr\Phi\rangle=\sum_{j=1}^r m_{I_j}\ .
}
In order to compare with our earlier results we can force the trace
part of $\Phi$ to vanish by considering, as previously, 
a more general potential $W(x)=\mu
x^2/2-\xi x$. The extra parameter $\xi$ acts as a Lagrange multiplier
which will be fixed by varying the effective superpotential. 

The second question that confronts us is how the one cut solution can
describe the multiplicity of vacua that we described
in Section \ref{Semi-Classical Analysis}.
The origin for the multiplicity of vacua is solved by realizing that
there is an ambiguity as to which sheet the points $m_I$ lie. It is
useful at this point to compare to brane configurations
in Section \ref{brane config}.
If $m_I$ lies on the bottom sheet then we interpret this to mean in
brane language that the corresponding semi-infinite D4-brane ends on
the right-hand NS5-brane, while, conversely, $m_I$ lies on the top
sheet then the semi-infinite D4-brane ends on the left-hand
NS5-brane. This means that the confining vacua, for example, have all
the $m_I$ on the bottom sheet.

For the one-cut solution, $y^2=(\mu x-\xi)^2+b_1$ and 
the glueball superpotential is
\EQ{
W_{\rm gb}=
\frac{\pi i\tau}{4\mu}b_1-\frac{1}{2}\sum_{I=1}^{2N}
\Big(\int_{m_I}^{e_1}y\,dx-W(m_I)\Big)\ ,
\label{oci}
}
where $e_1$ is a point at the end of the cut. 
We can evaluate the integrals in 
this expression explicitly to give
\EQ{
W_{\rm gb}=
\frac{\pi i\tau}{4\mu}b_1
+\frac{1}{4\mu}\sum_{I=1}^{2N}
\Big(\mu^2m_I^2-2\xi m_I+\mu \tilde m_I f_{\pm}(\tilde m_I)+b_1\log
\big(\mu \tilde m_I+f_{\pm}(\tilde m_I)\big)-\tfrac12b_1\log(-b_1)\Big)
\label{ep integral3}
}
where we have defined $\tilde m_I=m_I-\xi/\mu$. Note that
$f_{\pm}(x)=\pm\sqrt{\mu^2 x^2+b_1}$ depending upon whether $x$
is on the top or bottom sheet, respectively.

The critical point equation is
\EQ{
\prod_{I=1}^{2N}\big(\mu \tilde m_I+f_{\pm}(\tilde m_I)
\big)=(-b_1)^{N}e^{\pi i\tau}\ .
\label{cpe}
}
the critical value of the glueball superpotential is then
\EQ{
W_\text{gb}^*=\frac{1}{4\mu}\sum_{I=1}^{2N}\big(\mu^2
m_I^2-2\xi m_I+\mu \tilde{m}_I f_{\pm}(\tilde m_I)\big)\ .
}

\subsection{The vacuum structure}

In Section \ref{Semi-Classical Analysis}, we labelled the vacua by a subset 
$\{m_{I_1},\ldots,m_{I_r}\}\subset\{m_1,\ldots,m_{2N}\}$. 
We will identify this vacuum with the matrix model saddle-point solution
where $\{m_{I_1},\ldots,m_{I_r}\}$ are on the top sheet and the remainder
$m_I$, $I\not\subset\{I_1,\ldots,I_r\}$, are
on the bottom sheet. In order to see why this is the correct
identification, consider the classical limit $\tau\to i\infty$. In
this limit $b_1$ is small and we can find a solution to \eqref{cpe}
order-by-order in $b_1$. To leading order we find
\EQ{
(-b_1)^{N-r}=(2\mu)^{2(N-r)}e^{\pi
  i\tau}\frac{\prod_{I=1}^{2N}\tilde m_I}{\prod_{j=1}^r\tilde 
m_{I_j}^2}+\cdots\ .
}
and
\EQ{
W_\text{gb}^*=\sum_{j=1}^r\big(\frac\mu2m_{I_j}^2-m_{I_j}\xi\big)
-\frac{N-r}{2\mu}\xi^2
-\mu(N-r)\left(-e^{\pi
  i\tau}\frac{\prod_{I=1}^{2N}\tilde m_I}
{\prod_{j=1}^r\tilde m_{I_j}^2}\right)^{1/(N-r)}+\cdots\ .
}
The parameter $\xi$ is determined from $\partial
W^*_\text{gb}/\partial\xi=0$, giving to leading order
\EQ{
\xi=-\frac{\mu}{N-r}\sum_{j=1}^rm_{I_j}+\cdots\ .
}

For example, for the confining vacuum $r=0$ and all the $m_I$ are on
the bottom sheet and to leading order $\xi=0$. In this case, to leading order
\EQ{
W^*_\text{gb}=-\mu N\Big(-e^{\pi
  i\tau}\prod_{I=1}^{2N}m_I\Big)^{1/N}+\cdots\ ,
\label{mmcc}
}
which is the characteristic expansion in terms of fractional instantons.
Correspondingly, in a Higgs
vacuum, $r=N-1$, so $N-1$ of the $m_I$ are on the top sheet. In this
case, $\xi=-\mu\sum_{j=1}^{N-1}m_{I_j}+\cdots$ and 
\EQ{
W^*_\text{gb}=\frac\mu2\Big(\sum_{j=1}^{N-1}m_{I_j}^2+\big(
\sum_{j=1}^{N-1}m_{I_j}\big)^2\Big) 
+\mu
\frac{\prod_{I=1}^{2N}(m_I+\sum_{j=1}^{N-1}m_{I_j})}
{\prod_{j=1}^{N-1}(m_{I_j}+\sum_{j=1}^{N-1}m_{I_j})^2}e^{\pi i\tau}+\cdots
\ .
\label{mmhi}
}

Using the relation between the couplings in \eqref{mmh}, one 
can confirm that \eqref{mmhi} and \eqref{mmcc} for $N=2$ match precisely with
the results from the integrable system, \eqref{inth} and \eqref{intc},
respectively.  

\section*{Acknowledgements}

We would like to thank Nick Dorey for useful discussions.
KO also thank H.~Fuji, K.~Okunishi and T.~
Yokono. KO is
supported in part by PPARC for Research into Applications of Quantum
Theory to Problems in Fundamental Physics.

\end{document}